# Tamm Plasmon Resonance as Optical Fingerprint of Silver/Bacteria Interaction

Simone Normani[1], Pietro Bertolotti[1,2], Francesco Bisio[3], Michele Magnozzi[4], Francesco Federico Carboni[1], Samuele Filattiera[1], Sara Perotto[1], Fabio Marangi[1,2], Guglielmo Lanzani[1,5], Francesco Scotognella[5] and Giuseppe Maria Paternò[1,5]*

[1]Center for Nano Science and Technology@PoliMi, Istituto Italiano di Tecnologia, Via Giovanni Pascoli, 70/3, 20133 Milano, Italy
[2]Biomedical Engineering Department, Politecnico di Milano, Piazza Leonardo Da Vinci, 32, 20133 Milano, Italy
[3]SuPerconducting and other INnovative materials and devices institute (SPIN), Consiglio Nazionale delle Ricerche (CNR), Corso F.M. Perrone 24, 16152 Genova, Italy
[4]Dipartimento di Fisica, Università di Genova, via Dodecaneso 33, 16146 Genova, Italy
[5]Physics Department, Politecnico di Milano, Piazza Leonardo Da Vinci, 32, 20133 Milano, Italy
*Authors to whom correspondence should be addressed: giuseppemaria.paterno@polimi.it

## ABSTRACT

Incorporation of responsive elements into photonic crystals is an effective strategy for building up active optical components to be used as sensors, actuators and modulators. In these regards, Tamm Plasmon (TP) modes have arisen recently as powerful optical tools for the manipulation of light-matter interaction and for building sensors/actuators. These emerge at the interface between a dielectric mirror and a plasmonic layer and, interestingly, can be excited at normal incidence angle with relatively high quality factors. Although its field is located at the interface between the dielectric mirror and the metal, recent studies have demonstrated that corrugation at the nanoscale permits to access the TP mode from the outside, opening new exciting perspectives for many real-life applications.

Here, we show that the TP resonance obtained by capping a distributed Bragg reflector with a nanostructured layer of silver is sensitive to the presence of bacteria. We observed that nanoscale corrugation is essential for accessing the TP field, while the well-known bio-responsivity of silver nanostructures renders such a localised mode sensible to the presence of *Escherichia Coli*. Electrodoping experiments confirm the pivotal role of nanostructuration, as well as strengthening our hypothesis that the modifications of the TP mode upon exposure to bacteria are related to the accumulation of negative charge due to the bacterial-driven removal of $Ag^+$ ions from its lattice. Finally, we devised a case study in which we disentangled optically the presence of proliferative and non-proliferative bacteria using the TP resonance as a read-out, thus making these devices as promising simple all-optical probes for bacterial metabolic activity, including their response against drugs and antibiotics.

**INTRODUCTION**

Distributed Bragg reflectors (DBRs) are simple multi-layered systems in which a photonic band gap (PBG) arises from the periodic arrangement of materials with different refractive indices, in analogy with the well-known phenomenon of x- ray reflection. Although DBRs have been mostly employed for fundamental studies and/or for building up passive optical components, in the last couple of decades the introduction of selected responsive elements has paved the way to the effective utilization of DBRs as simple yet effective tools for the active manipulation of light and for sensing purposes.[1-3]

Recently, plasmonic materials have emerged as intriguing responsive elements in hybrid plasmonic/photonic devices.[3-8] Beside the incorporation of plasmonic nanostructures as integral part of the DBR, in which the modification of the plasmon frequency and the complex dielectric function via electro/photodoping permits an active manipulation of the PBG,[3,6,7,9-13] another approach exploits hybrid plasmonic/photonic optical modes at the interface between metal and dielectric materials. Within this framework, we have recently incorporate a thin layer of silver on top of DBRs for the purpose of bacterial detection.[14,15] Briefly, our idea was to exploit the well-known capability of silver to interact with bacteria, to a change the refractive index conditions at the interface between our hybrid plasmonic/photonic system and air. However, in those proof-of-concept studies our information was encoded on the shift of the PBG, whose broadness (full width half maximum, FWHM, exceeding largely 100 nm) rendered the read-out difficult to be interpreted. To circumvent such an issue, localised defective cavity modes are taken usually as reliable spectral signature for the presence of various analytes, including bacteria.[16] Nevertheless, such modes are often buried inside the photonic lattice, rendering field accessibility a not trivial problem.

Alternatively, localised defective modes can emerge at the interface between dielectric mirrors and plasmonic metals. This is the case of the so-called Tamm plasmons (TPs), which are electromagnetic modes confined between a DBR and a noble metal layer (*i.e.* gold and silver).[14-16] In contrast to surface plasmons, TPs can be excited with high quality factor at normal incidence angle, a feature that is very appealing for the modulation/exploitation of light-matter interaction.[20,21] Beside this advantageous operative feature, the relative simplicity of the planar structure necessary to achieve the TP resonance lends itself to both facile fabrication procedure (*i.e.* spin-casting and thermal evaporation) and large-scale fabrication. So far, TPs have been utilized for many purposes, including lasing,[20] modification of light-matter interaction,[21,22] thermal emitters[23] and optical sensors,[20-24] among others. However, this comes with a disadvantage, as the electric-field distribution of the TP mode is located predominantly at the DBR/metal interface, thus being almost inaccessible from the outside. This would limit greatly the operational possibilities of such devices, *i.e.* for sensing applications. On the other hand, it has been demonstrated recently that patterning or corrugation of the metal film at the micro/nanometre level enables both reduction of metal losses and exposition of the TP field to external stimuli, such as changes of the refractive index.[25-27]

Here, we observe that the TP mode originating at the interface between a nanostructured layer of silver and a DBR is highly responsive to exposure to bacteria, as highlighted by a blue shift and a clear damping of the TP resonance as compared with the control measurements. Our data indicate that such a modification of the TP-mode can be ascribed to the high bioactivity of silver, leading to a change of Ag charge carrier density owing to the silver ions uptake by bacteria. Electrodoping experiments suggest that corrugation at the nanoscale plays a significant role in the detection mechanism, as in this case the TP field enhancement is exposed to the air interface. To take advantage of these features, we devised a proof-of-concept experiment, in which we tested the capability of such TP devices to discriminate between proliferative and non-proliferative bacteria. Intriguingly, we cannot observe significant bacterial-induced changes on the TP resonance when we rendered bacteria non-proliferative via kanamycin or chloramphenicol administration. This can hold promise for the application of TP devices as simple optical drug screening platforms.

## RESULTS

**Device morphology and optical properties**. A sketch of the DBR and its cross section are presented in **Fig. 1a,b**, respectively: it simply consists of 4 bilayers of silica/titania deposited via spin-coating from their colloidal nanoparticle suspensions, and a 20 nm layer of Ag deposited on top of the DBR via thermal evaporation. Such a thickness ensures the emergence of the TP mode, while permitting to characterise the device also via transmission measurements (**Fig. S1**). The transmission/reflection spectrum reports clearly a PBG located at 550 nm - 750 nm (**Fig. 1c**), while the addition of the Ag layer increases the overall reflectivity of the samples across the visible range and leads to the rise of a relatively narrow transmission peak (and a corresponding drop in reflection) positioned close to the PBG low-energy edge, at around 725 nm. This peak can be ascribed to the emergence of the TP resonance.[24] Scanning electron microscopy images on the capping Ag layer (Fig. 1d) reveal the presence of nanoplates, with irregular shapes and sizes (average radius = 42 nm, average thickness = 20 nm, **Fig. S2**) in analogy with what reported in previous experiments.[28-30] The nanostructured Ag layer has an important role here for three important reasons: *i.* it leads to the appearance of the TP spectral feature that is narrower than the PBG (FWHM = 27 nm *vs*. 138 nm), whose intensity and position thus being a more sensible read-out upon exposure of analytes; *ii.* the intrinsic and high bio-responsivity of nano-silver, which results into bacteria-driven modifications of silver plasmonic properties[31,32]; *iii.* the corrugation at the nanoscale that does not hamper the emergence of the TP resonance, while allowing a direct access to its field. Hence, our main idea is to exploit all these three advantages to build up an optical sensor capable to detect the presence of bacteria, as well being able to map out their metabolic activity, with the broad view to develop drug-testing platforms.

Exposure of the device to *E. coli* colonies in an Agar plate leads to the formation of extended bacterial communities on top of the surface (Fig. 1d), with the bright spots located in the close proximity of the bacterial membrane that can be linked to the uptake and incorporation of Ag$^+$ ions clusters inside the cells. Such a mechanism is commonly recognised as the main driving force of nano-silver/bacteria cytotoxicity.[33] The presence of a well-defined zone of inhibition in the contaminated Agar medium can linked to the eradication of bacteria, confirming the biocidal activity of the Ag layer (**Fig. S3**).

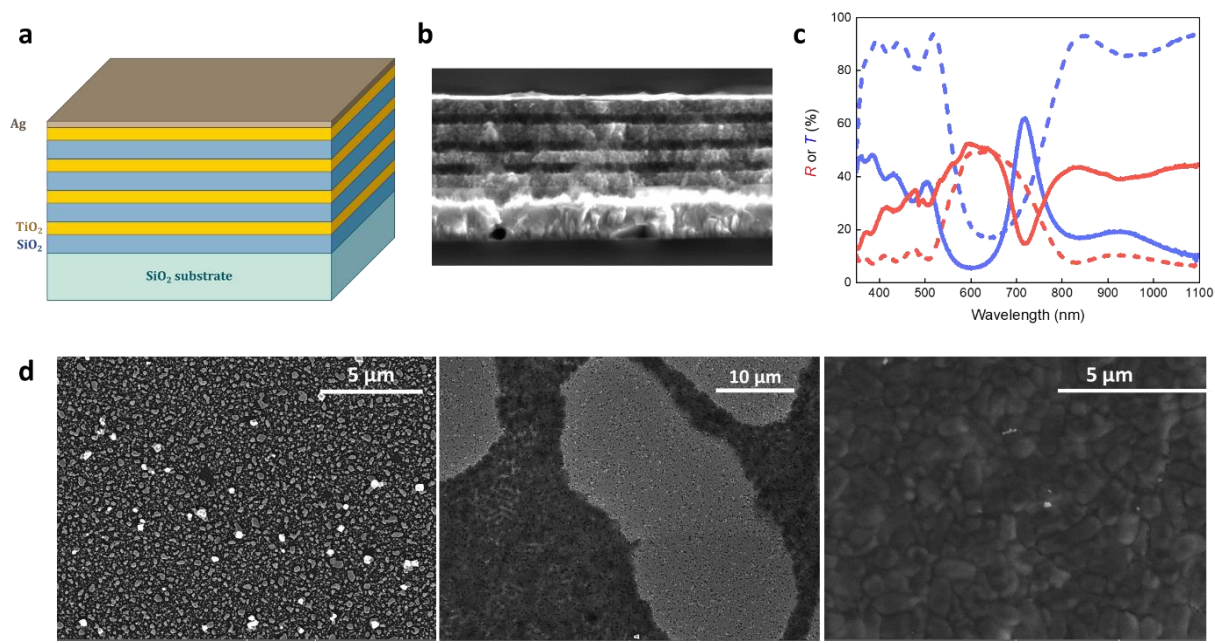

**Figure 1. Structure, morphology and optical read out of the TP device. a.** Sketch of the DBR with the Ag capping layer. **b.** SEM cross section of the multi-layered system. We worked out a layer thickness of 130/80 nm for silica/titania, and a capping layer of about 20 nm. Optical simulations reproduced with a good degree of approximation the spectral feature of the such device, with thickness of 110/70 nm and a capping layer of 30 nm (see materials and method and supplementary information section). **c.** Reflection and transmission spectra of a typical sample, before (dashed lines) and after (solid lines) Ag deposition, showing the rise of the Tamm plasmon resonance around 725 nm. **d.** SEM images of the Ag layer before (left panel), after *E. coli* exposure (central panel) and zoom on the bacterial colonies to highlight the formation of Ag cluster inside the cells.

**Exposure to *E. coli* modifies the TP resonance**. We then proceed to the evaluation of the effects of bacterial exposure on the plasmonic/photonic spectral response. With this in mind, we first checked how a bare multilayer without silver behaves upon exposure to the culture medium (Luria-Bertani broth, control sample), and to *E. coli* cells (**Fig. 2a**). In this case, we observe a broadening of the PBG owing to the infiltration of the medium in the porous structure, without any clear difference between the samples exposed to LB and bacteria. After having assessed that the bare DBRs cannot detect clearly the presence of bacteria, we then passed to the evaluation of TP devices. Here, we see strong modifications of the spectral features upon contamination with bacteria (**Fig. 2b**) when compared to the LB treated sample namely: *i.* a broadening of the PBG (FWHM = 160 nm *vs.* 128 nm, for *E. coli* and LB respectively) that we attribute to the inhomogeneous population of bacteria on top of the silver layer, which increase the degree of optical disorder (refractive index) at the interface with air; *ii.* a clear blue shift (22 nm) of the TP and a damping of its transmission (20 %). To highlight these observations, we plotted the differential transmission spectra (ΔT/T) for LB and *E.* coli, which is essentially the difference between the spectrum of the perturbed sample (bacteria) and the control sample (LB), normalised to the transmission of the control (dotted line in **Fig. 2c**). Specifically, the negative signal centred at around 760 nm accounts for both Tamm mode blue-shift and decrease of transmission, thus it can be taken as a diagnostic spectral parameter for identifying the presence of bacteria. Note that the blue shift and decreased transmission observed here have the same physical origin of the spectral changes observed recently in the plasmon resonance of nanostructured silver layers upon contamination with *E. coli*.[31] By combining ultrafast optical spectroscopy, x-rays diffraction and imaging, we assigned those changes to the silver oxidative dissolution in presence of bacteria.[32] In this work, we enhance the detection sensitivity by transducing the bacterial effect into a relatively narrow TP resonance. To be quantitative, the blue shift of the silver plasmon resonance upon bacteria exposure divided by its FWHM lies around 5%,[31] while for our previous silver/DBR devices (without Tamm resonance) the PBG blue-shift/FWHM ratio amounts to 7%.[14] On the other hand, interestingly, the blue shift/FWHM ratio for the Tamm resonance is more than one order of magnitude higher than in the abovementioned cases (80%), suggesting that this approach can afford a relatively high sensing capability.

We also assessed the capability of our hybrid device to monitor the presence of a different bacterium, namely *Salmonella enterica*. In analogy with *E. coli*, this is a Gram-negative bacterium of pharmacological and medical interest due to its pathogenicity. Again, we observe a ∼ 20 nm blue shift of the TP mode (**Fig. S4**), suggesting that our observations are somehow general, at least for bacteria sharing the same cellular envelope.

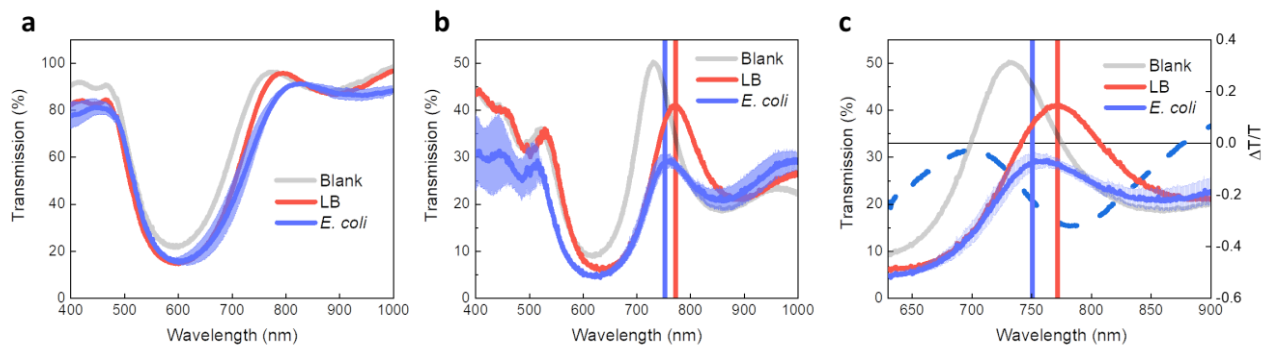

**Figure 2. The effect of bacterial exposure on TP-devices. a.** Transmission spectra of pristine silica/titania multilayers (blank) and exposed to LB and *E. coli* cells. We take both transmission and reflection measurements, while for simplicity sake we show only transmission data in the main text. All the measurements involving bacterial cells were taken on biological triplicates. The curve represents the average over those measurements and the shadow is the standard deviation. **b.** Transmission spectra of TP-devices exposed to LB and *E. coli*. **c.** Zoom on the TP-resonance

spectral region, highlighting the 22 nm blue shift and decrease of transmission of the TP state upon contamination with bacteria. The dashed line represents the differential spectrum (ΔT/T) calculated as $(T_{E.coli} - T_{LB})/T_{LB}$), which highlight the modification of the spectral response (both shifts and changes in transmission) of the TP resonance when exposed to bacteria.

**Metal corrugation affects TP response to exogenous stimuli.** To evaluate the role of the nanoscale corrugation of our films on the accessibility of the TP field, we fabricated and characterised TP devices with gold as a metallic plasmonic layer. We selected gold owing to its biocidal activity,[34] while ensuring a well-defined TP resonance.[24] This metal layer (20 nm) appears as a compact film, in contrary to what observed in silver (**Fig. 3a**). We thus proceeded to the evaluation of the bio responsivity of such TP device by exposing it to *E. coli*, following the same protocol adopted for silver-based systems. In this case, the blue shift was clearly less prominent than in the silver-based TP (5 nm), while we could not note any substantial decrease of transmission. We reckon that this can be attributed to the less corrugated Au layer than Ag, which in turns can drive the following effects: *i*. decrease of bioactivity due to the relatively small surface/volume ratio in smooth layers. Metal structuration at the nanoscale is one of the most important feature necessary for achieving effective biocidal activity, as the surface available for ionic release (oxidative dissolution) and bacterial uptake decrease considerably passing from nanostructured to bulky materials ;[35] *ii*. inaccessibility of the TP field from the outside.[28]

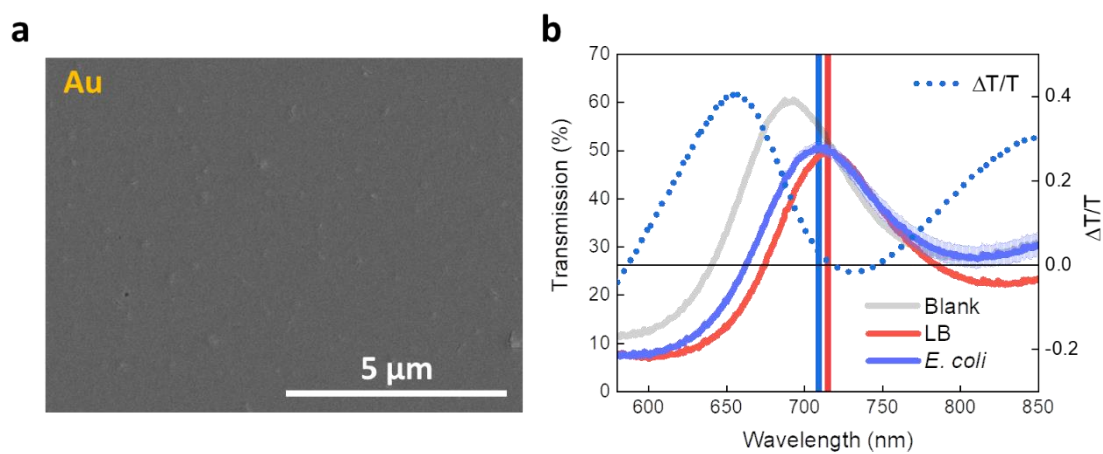

**Figure 3. The effect of bacterial exposure on gold-based TP devices. a.** SEM image of the Au layer (thickness 20 nm, obtained via thermal evaporation). **b.** TP-resonance of the device exposed to *E. coli* cells. Also in this case, we report the differential transmission to highlight the difference between the perturbed sample and the control (dashed line).

We then proceed to an additional experiment, aiming at disentangling these two phenomena concurring to the less effectiveness of gold-based TP than silver. To this end, we set-up an electrodoping experiment, by taking inspiration from our recent works on electro-responsivity of DBRs.[7,9,10] Specifically, we fabricated our TP device on an Indium Tin Oxide (ITO) electrode and sandwiched with a second ITO layer on top, to provide electrical connection (**Fig. S5**). We then applied a negative or positive DC voltage to the composite structure and looked at the transmission spectrum in real time. The transmission spectra show a clear shift of the TP peak as a function of applied voltage: the peak undergoes a red shift (11 nm) and increase of transmission under positive bias, while we observe a blue shift (8 nm) and a decrease of transmission under negative bias (**Fig 4a**). As briefly mentioned above, we proposed that the bacterial-driven modification of the silver plasmon resides on the release of positive metal ions. This can be essentially rationalised as a "biological doping",[14] as removal of $Ag^+$ will leave behind an excess of negative charge. The resulted plotted in figure 4 are thus fully consistent with this interpretation, for both phenomena leads to an accumulation of negative charges in the plasmonic metal. In addition, these results confirm that the TP field can be indeed accessed from the outside, provided that metal is corrugate/structured.

We verified this by performing the same experiment on smooth gold-based devices (**Fig. 4b**). Here, we could not observe any significant effect on TP peak position, while we noted a small decrease/increase of transmission (2 %) upon application of a negative/positive bias. To highlight the higher magnitude of electric-responsivity of the silver TP than the gold-based resonance, we plotted the differential transmission of these two systems (**Fig. 4c**). These results confirm that metal corrugation essentially dictates accessibility of the TP field from the outside, rendering this resonance highly appealing for building up sensors, actuators and modulators. Interestingly, by means of optical modelling, we were able to reproduce both the occurrence of the TP mode and its blue shift upon increase of electron density, hence corroborating the experimental data. All the details about modelling and simulated spectra can be found in the supplementary information section and in **Fig. S6**.

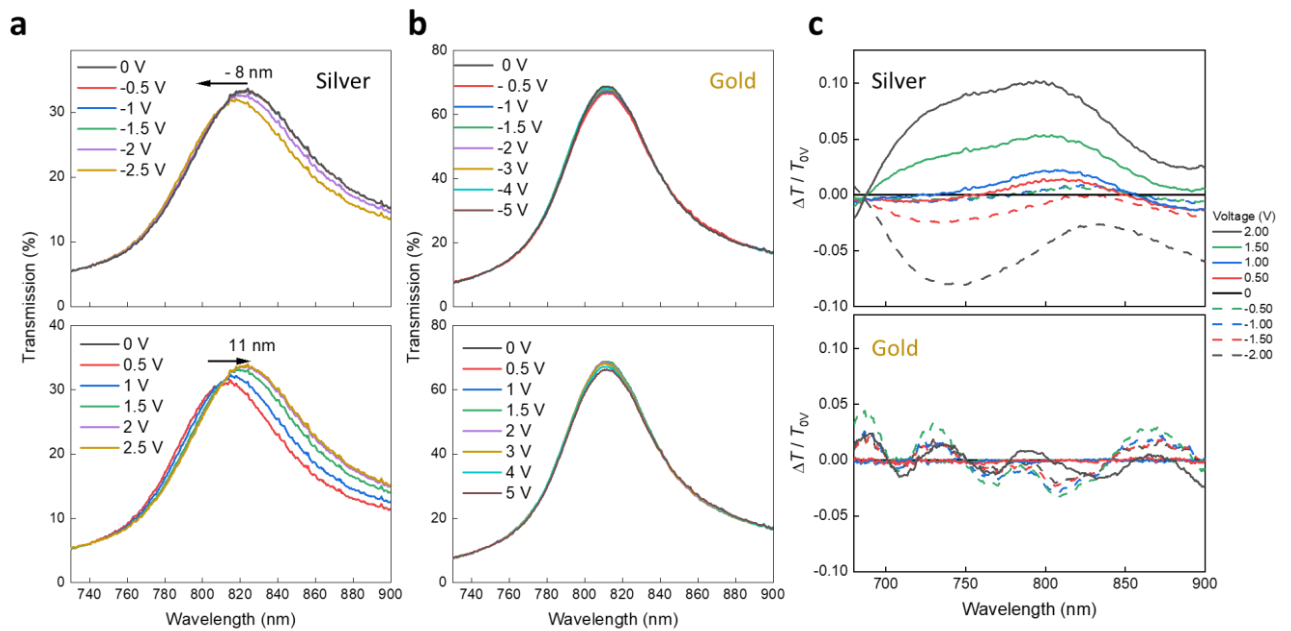

**Figure 4. Electro-responsivity of Ag- and Au-based TP resonance. a.** Transmission of the TP resonance in silver-based device upon application of negative (top) and positive (bottom) voltage. **b**. Transmission of the TP resonance in gold-based device upon application of negative (top) and positive (bottom) voltage. **c**. Differential transmission spectra upon application of positive and negative bias for silver- (top) and gold-based TP devices.

**TP resonance probes proliferative status of bacteria.** Since the uptake of $Ag^+$ ions by bacteria is driven by their metabolic processes, for instance those controlling expression of nucleophiles-containing proteins (*i.e.* thiols and amino groups),[36,37] one would expect a relationship between the modulation of TP resonance and their proliferative status. In general, the link between metabolic activity and antibiotic/drug bacterial uptake has been investigated deeply,[38] while in our case it would allow building up optical sensors for bacterial metabolic activity, including their response against external stressors (*i.e.* drugs and antibiotics), which rely on the easy-to-assess TP read-out. To verify such hypothesis, we then devised a proof-of-concept experiment in which we exposed the devices to both proliferative and non-proliferative *E. coli* cells. This latter biological sample was prepared via the administration of two selected drugs: the antibiotic kanamycin and the bacteriostatic drug chloramphenicol. The former inhibits protein synthesis by compromising ribosomes activity, leading eventually to a complete lack of protein turnover,[39] while the latter permits to achieve the same result by preventing the elongation of peptide chain on the 50S subunit of ribosomes.[40] It is thus expected that, in this case, interaction between bacteria and silver cannot occur effectively, due to the excessively low viability of the cell population. Indeed, we observed that non-proliferative bacteria leads to only 5 nm blue shift and virtually no decrease of the transmission intensity of the TP resonance (**Fig. 5a**). On the other hand proliferative *E. coli* cells leads to ~20 nm blue shift and 20% damping of the transmission intensity, in analogy with what observed in the previous sample batches. These effects are highlighted in the differential spectra (**Fig. 5b**), in which the diagnostic negative signal at 760 nm due TP blue shift and decrease of transmission results greatly reduced for the device exposed to non-proliferative bacteria.

We then prepared another device batch to evaluate the effect of the chloramphenicol-treated bacteria on the Tamm mode (**Fig. 5c**). Here, we observe that the Tamm resonance overlaps to a lesser extent with the control curve if compared with the case of Kanamycin, indicating that such drug is likely less effective than Kanamycin in inhibiting the capability of silver to interact with bacteria. This is also underlined by the diagnostic negative signal lying at around 760 nm (**Fig. 5d**), which is appreciably less damped than in the case of Kanamycin. Based on such differential curves, one can conceive a simple ratiometric sensing approach, in which the differential transmission at 760 nm measured with a stable light source (*i.e.* laser) can be taken as an indicator of the proliferative activity of bacteria and, thus, of their susceptibility against external stressors.

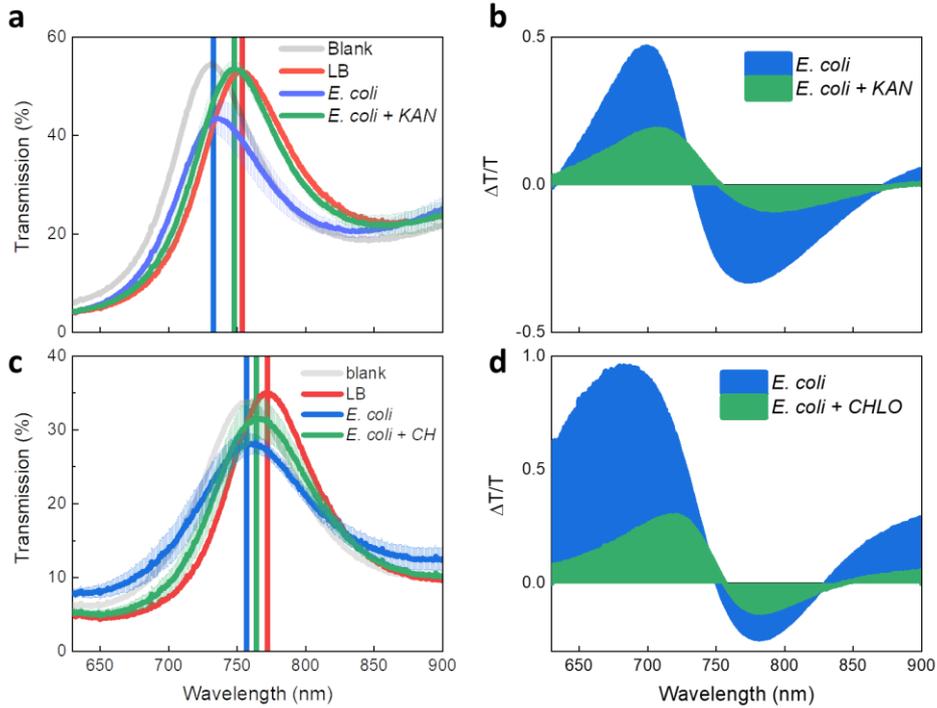

**Figure 5. TP resonance is sensible to the proliferative status of bacteria. a.** TP-resonance of the device exposed to *E. coli* and *E. coli* treated with Kanamycin (KAN, for brevity). **b.** Differential transmission spectra (normalised to the transmission of the control measurement, LB), highlighting the modification of the TP resonance upon exposure to *E. coli* and *E. coli* treated with Kanamycin. **c**. TP-resonance of the device exposed to *E. coli* and *E. coli* treated with Chloramphenicol (CH for brevity). **d.** Differential transmission spectra (normalised to the transmission of the control measurement, LB), highlighting the modification of the TP resonance upon exposure to *E. coli* and *E. coli* treated with Chloramphenicol.

## CONCLUSIONS

In this paper, we have demonstrated that TP are sensible to both the presence of bacteria and, interestingly, to their proliferative status. Specifically, we fabricated DBRs via facile spin-casting deposition from colloidal dispersion of silica/titania nanoparticles, and capped the dielectric mirror with a nanostructured layer of silver. The corrugation at the nanoscale did not hamper the occurrence of a well-defined TP resonance at the low-energy side of the PBG, in-fact it permitted to access the TP field from the outside, as revealed by our electro-doping experiments. The modification of the silver optical properties brought about by the model bacterium *E. coli*, which are essentially linked to bacterial-driven oxidative dissolution of the metal, are translated into the modulation of TP resonance mode intensity (decrease of transmission) and spectral position (blue shift). Electrodoping experiments and theoretical modelling confirmed that such effects could be connected to an excess of electron density in the metal layer, owing to the removal of positive ions from its lattice. Finally, as a case study we tested the capability of our TP device to discriminate between proliferative and non-proliferative bacteria. Taken together, these data can pave the way to the introduction of optical sensors whose simple read out can be exploited for monitoring the metabolic status of bacteria, with the view to build up drug testing platforms.

## METHODS

**Fabrication.** Silica ($SiO_2$) and titania ($TiO_2$) were the materials chosen for the fabrication of the colloidal photonic crystals,[41] due to their high transparency in the visible range, their relatively high refractive index mismatch, and their availability. In order to better match in wavelength as much as possible the localized surface plasmon resonance (LSPR) of silver, set in the near infrared (NIR) range, the DBRs were designed to have a PBG extending into the red visible range (beyond 600 nm). While this is by no means a necessity in order to obtain a TP, it was nevertheless observed that reducing the mismatch between the PBG and the LSPR wavelengths produced a narrower TP peak, and therefore better resolution in measuring its spectral shifts. In order to centre the PBG around 650-700 nm, keeping into account an expected porosity of the layers around 20 -30%, the thickness of the dielectric layers should be about 110-130 nm for $SiO_2$, and 70-80 nm for $TiO_2$. These values were estimated using a DBR simulation program based on the transfer matrix formalism and using the Maxwell Garnett model for glass layer porosity.[42] The DBR were then fabricated by spin-coating: a 30 wt% aqueous solution of $SiO_2$ nanoparticles (*Ludox*) was diluted with distilled water to a 6 wt% concentration, and a $TiO_2$ nanoparticle powder (*GetNanoMaterials*) was dissolved in distilled water to prepare a 10 wt% aqueous solution. These concentrations were estimated in order to obtain the required thickness of $SiO_2$ and $TiO_2$. The two solutions were then processed via a tip sonicator (*Branson 450 Digital Sonifier*) to homogenize them and break apart particle clusters, and then filtered with 0.22 μm filters to remove remaining clusters. The DBR were then prepared by depositing four pairs of alternating $SiO_2$ and $TiO_2$ layers on 2 cm squares of microscope glass, dropping 150 μl of solution per layer, spinning the sample at 2000 rpm with a spin-coater for 1 minute, and annealing it on a hot plate at 350° C for 20 minutes after every layer deposition. Finally, a silver (gold) layer was deposited on top of the DBR by thermal evaporation (thickness 20 nm). Several layers of silver with different thicknesses (every 5 nm) were deposited in order to optimize the TP peak resolution, eventually determining via spectroscopy measurements that the optimal silver thickness is around 20 nm, for which the TP peak is deepest and narrowest. Electrodoping experiments were carried out following the same experimental procedure used in our previous work.[7]

**Optical spectroscopy and scanning electron microscopy.** The visible and near infrared (VIS-NIR) transmission and reflection measurements were performed on the samples before and after exposure to the bacterial cultures, using a deuterium-halogen lamp (*AvaLight-D(H)-S*) and a fiber-coupled spectrometer (*Avantes, AvaSpec-HS2048XL-EVO*), averaging over 30 measurements with an integration time of 2 ms. The measured data were normalized to the spectrum of a reference blank sample (silica glass slide for transmission, silver mirror for reflection) and a dark baseline. Reflection measurements in particular were averaged over 5 different spots across each sample.

Imaging of the sample surfaces was performed using scanning electron microscopy (SEM , *Tescan*) with various degrees of magnification, so as to assess the presence of *E. coli* and the state of the bacterial cells in contact with both plain glass and silver-coated surfaces.

***E. coli* cultures and sample exposure.** Micro-organisms used for experiments are from the 25922 strain of *Escherichia coli* and 14028 strain of *Salmonella enterica* (provided by *ATCC*). Luria-Bertani (LB) broth and LB agar were used respectively for liquid culture and on-plate growth assays. The liquid bacterial cultures were grown overnight in LB medium in an incubator at a constant temperature of 37 °C, with a 200 rpm agitation rate. They were then diluted to $OD_{600}$ 0.5 before moving them to LB agar plates. Each sensor was then exposed on an agar plate to a volume of 100µl of liquid culture, and left at 37 °C overnight. For sensor tests involving antibiotics, cultures were diluted to $OD_{600}$ 0.5, and later exposed to kanamycin A ($C_{18}H_{36}N_4O_{11}$, 50 µg/ml in water) or chloramphenicol ($C_{11}H_{12}N_2O_5Cl_2$, 170 µg/ml in ethanol) for 12 hours.

**Optical simulations.** The simulations were performed by means of the WVASE software (J.A. Woollam Inc.). The optical model consisted of a stack of dielectric layers, each characterized by its own thickness and complex dielectric function. Fresnel boundary conditions at the interface between the layers were assumed. For the simulation of the DBR backbone, the model (bottom to top) included: i) a silica substrate, ii) 4 identical repetitions of a nanoporous silica layer and a nanoporous titania layer. The nanoporous layers were modelled as Bruggeman effective media composed of a silica (titania) backbone and voids, as already proven appropriate for these systems.[43] The dielectric functions of the silica and the titania backbone were taken from ref.[44] and ref.[45], respectively. During the fit, the dielectric functions of the backbones were kept constant and only the thickness and the void fraction of the porous layers were left free to vary.

**AUTHOR'S CONTRIBUTIONS**

S.N. designed and fabricated the samples, performed optical measurements, and analysed the data. P.B. grew, exposed the samples to the bacterial cultures, and performed the antibiotics experiments. F.B. and M.M performed the optical simulations. F.F.C. fabricated the samples, performed optical measurements, and analysed the data. S.F. carried out the electrodoping experiments together with S.N.  S.P. and F.M. carried out the SEM measurements. G.L. conceived and supervised the work. F.S. supervised the work, and assisted in the data analysis. G.M.P. conceived and supervised the work, designed the experiments, and assisted in the data analysis. All authors contributed to manuscript drafting and revising.

## ACKNOWLEDGEMENTS

This work has been supported by Fondazione Cariplo, grant n° 2018-0979 and n° 2018-0505. F.S. thanks the European Research Council (ERC) under the European Union's Horizon 2020 research and innovation programme (grant agreement No. [816313]).

## CONFLICT OF INTERESTS

The authors declare no competing interests

## DATA AVAILABILITY

Original data are available upon request to the authors.

Simone Normani[1], Pietro Bertolotti[1,2], Francesco Bisio[3], Michele Magnozzi[4], Francesco Federico Carboni[1], Samuele Filattiera[1], Sara Perotto[1], Fabio Marangi[1,2], Guglielmo Lanzani[1,5], Francesco Scotognella[5] and Giuseppe Maria Paternò[1,5]*

[1]Center for Nano Science and Technology@PoliMi, Istituto Italiano di Tecnologia, Via Giovanni Pascoli, 70/3, 20133 Milano, Italy
[2]Biomedical Engineering Department, Politecnico di Milano, Piazza Leonardo Da Vinci, 32, 20133 Milano, Italy
[3]SuPerconducting and other INnovative materials and devices institute (SPIN), Consiglio Nazionale delle Ricerche (CNR), Corso F.M. Perrone 24, 16152 Genova, Italy
[4]Dipartimento di Fisica, Università di Genova, via Dodecaneso 33, 16146 Genova, Italy
[5]Physics Department, Politecnico di Milano, Piazza Leonardo Da Vinci, 32, 20133 Milano, Italy
*Authors to whom correspondence should be addressed: giuseppemaria.paterno@polimi.it

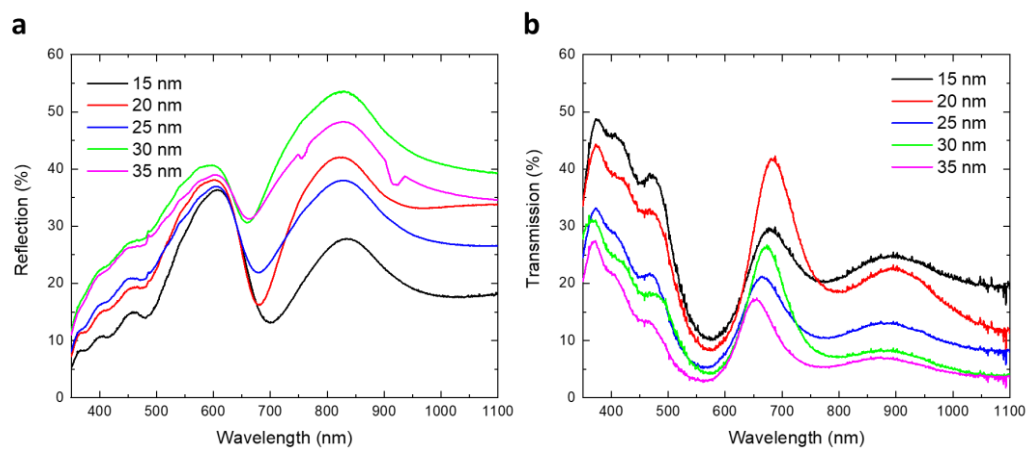

**Figure S6. Optimisation of the metal layer for the development of the TP resonance. a**. reflection and **b**. transmission of the TP device with different Ag layer thicknesses, keeping constant the DBR parameters.

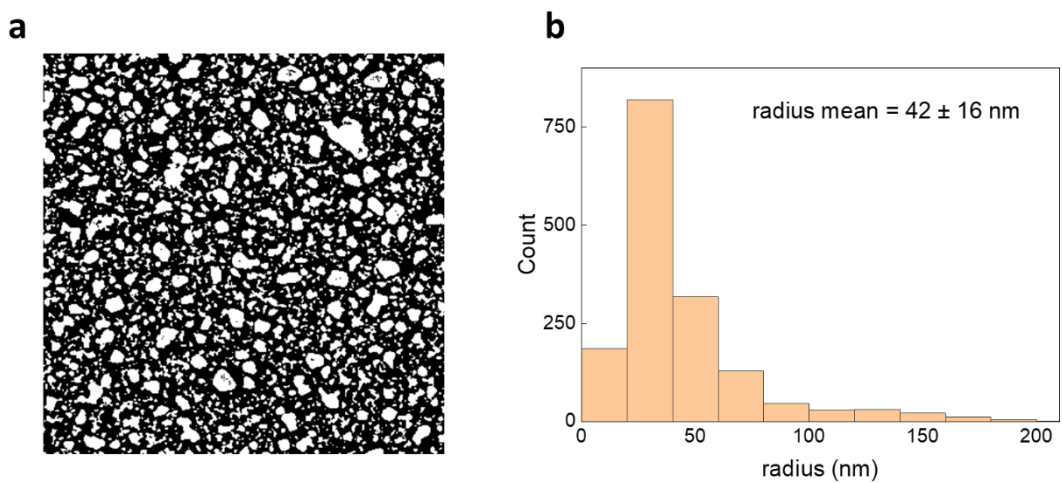

**Figure S7. Particle size analysis carried out via SEM. a.** representative region from which we extract the particle size distribution. **b.** Particle size distribution. The analysis was carried out by using the software ImageJ.

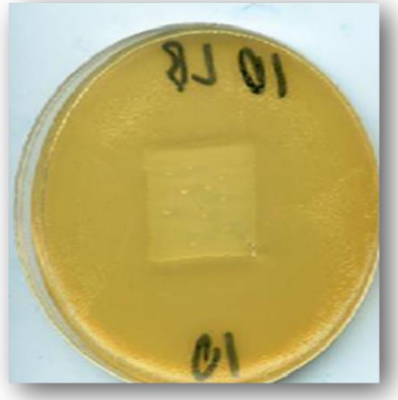

**Figure S8.** Zone of inhibition left behind from our TP device on a Agar Plate contaminated with *E. coli* cells. This confirms the biocidal activity and, thus, the bio-responsivity of the capping silver layer.

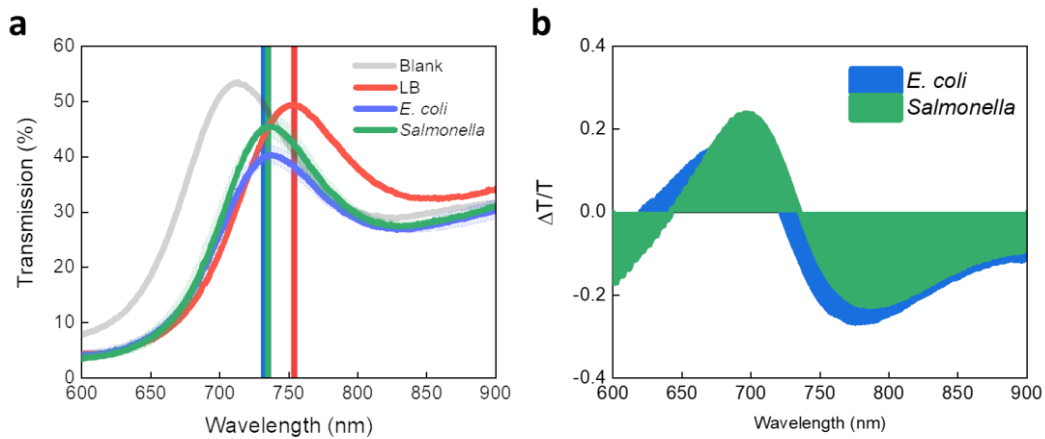

**Figure S9. The effect of *Salmonella enterica* exposure on TP-devices. a.** Transmission of the TP-resonance upon contamination with E. coli and Salmonella. **b**. differential spectrum (ΔT/T) calculated as $(T_{bacteria} - T_{LB})/T_{LB}$, which highlight the modification of the spectral response (both shifts and changes in transmission) of the TP resonance when exposed to bacteria. The two differential spectra almost overlaps, indicating that the TP resonance cannot discriminate between bacteria exhibiting the same cell wall (Gram-negative).

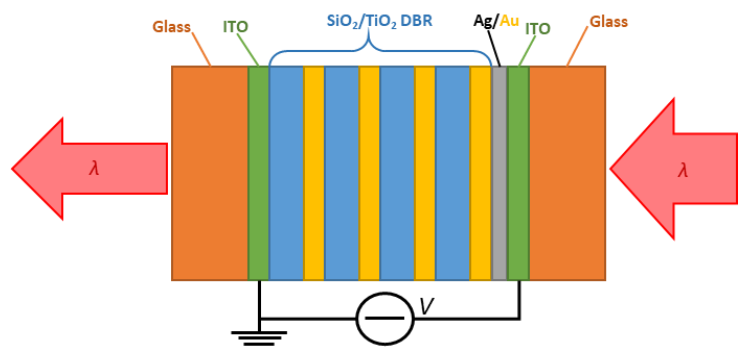

**Figure S10.** Sketch of the experimental apparatus used for electrodoping experiments. For these experiments, we followed virtually the same procedure that we employed in reference.[1]

**Optical simulation of the TP device.** In order to rationalize the experimental observations, we performed a set of simulations of the optical response of the DBRs (**Fig. S6**). The simulations were done for pristine DBRs, before and after the deposition of the Ag layer, and then the presence of the LB and the bacteria was effectively simulated.

In Figure S6, top panel, we report the experimental reflectance curve (green markers) of a DBR, along with the simulated curve corresponding to the best-fit morphological characteristics of the silica and titania porous layers. The best fit was achieved with a layer thickness of 110 nm (71 nm) and a void fraction of 24% (21%) for the silica (titania) layers. The model, correctly reproduces all the main spectral features of the reflectivity, with only minor intensity discrepancy. The optical response of the DBR with an additional Ag layer was modelled by adding a further dielectric layer on top of the previously-fitted DBR. The dielectric function of the discontinuous Ag layer on top of the DBR was modelled as a linear effective layer, i.e. a linear combination of the dielectric function of Ag (taken from reference[2]) and the one of the ambient (voids in this case). The layer thickness and the void fraction were left as free parameters. In Figure S6, bottom panel, we report the experimental reflectance (blue markers) along with the best fit (red line), corresponding to a film thickness of 33 nm and a void fraction just below 50 %. The appearance of the reflectance dip corresponding to the Tamm plasmon is correctly reproduced.

Modelling the optical response of the system following the interaction of the system with the bacteria requires some simplifications. We will examine the general trends as a function of individual variations of the system parameters that mimic the interaction with bacteria. The factors that may affect the optical response are: i) the immersion of the system in the LB broth, ii) the re-arrangement of the Ag layer and iii) the dispersion of Ag ions in the bacteria. The simulated curves are reported in Figure S6b. The black curve represents the best fit for the DBR+Ag system in air. The interaction with the LB broth can be reproduced considering the ambient to be a transparent medium with n=1.33 refractive index (including the voids in the Ag layer). Applying this change (red line in Figure S6b), we observe a general slight decrease of reflectance, and no significant spectral shifts of the Tamm plasmon.

The variation of morphology of Ag can be modelled by toggling the Ag fraction in the effective layer, while varying the film thickness to keep the effective amount of Ag constant. This is done under the assumption that the system is immersed in the LB. The effect of this modification on the Tamm plasmon is a red shift when the Ag content decreases (and the thickness correspondingly increases) and vice versa. To give an idea, if the Ag content in the effective layer decreases from 50% to 40%, and the thickness increases to 41 nm (to keep the total Ag constant), then the Tamm plasmon redshifts by 7 nm (compare green line with red line in Fig. S6b).

The dispersion of Ag ions in the bacteria can be represented by an increase in the free-electron density of Ag. From the optical point of view, if this change is moderate, it can be represented by a corresponding variation of the free-electron contribution to the dielectric function (the so-called Drude term). To give a quantitative estimation, augmenting the free-electron contribution by 5% led to a Tamm-plasmon blueshift of around 3 nm (compare blue line with red line in Fig. S6b). The simulations prove that there is a number of factors that affect the spectral characteristics of the Tamm plasmon. The experimental behavior recorded originates from a weighed superposition of these factors. The prevalence of redshifts over blueshifts in different experimental stages can be ascribed to the corresponding prevalence of each of the above-described factors on one another. We thus conclude that in our case, the accumulation of electron charge density would represent the most probable scenario.

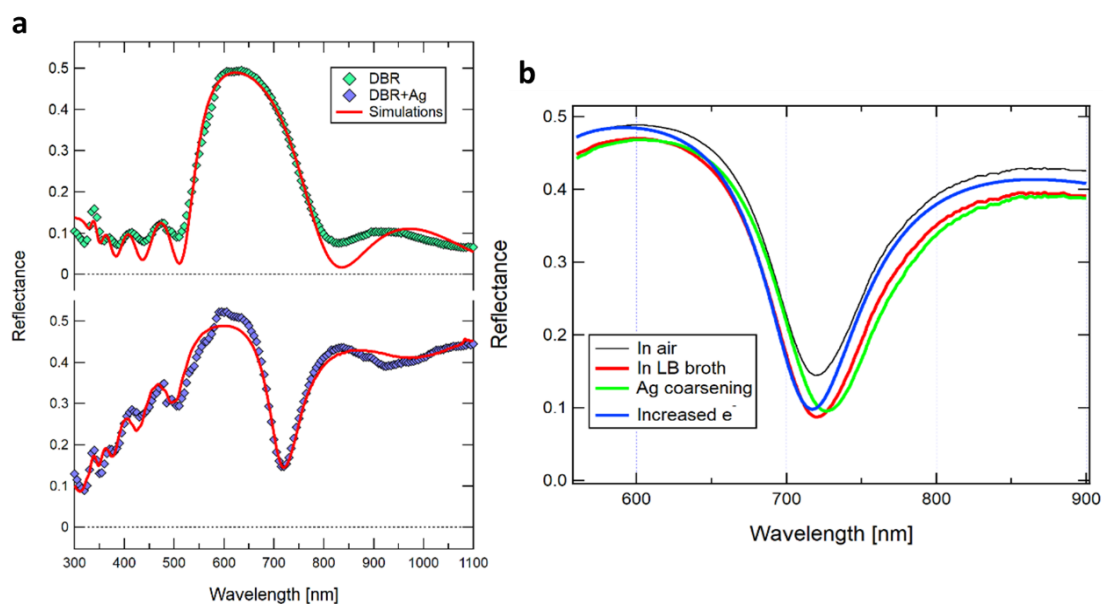

**Figure S11. Optical simulation of the TP device. a.** top panel: experimental reflectance of the DBR (green markers) and best fit (red line). Bottom panel: experimental reflectance of the DBR+Ag system (blue markers) along with its best fit (red line). **b**. Simulated optical response of the DBR+Ag upon interaction with bacteria. Black line: DBR+Ag in air.

Red line: DBR+Ag immersed in the LB broth. Green line: effect of Ag coarsening in LB broth. Blue line: effect of augmenting the free-carrier density in Ag, in LB broth (see text for details).